# Chemistry of layered *d*-metal pnictide oxides and their potential as candidates for new superconductors

Tadashi C. Ozawa*† and Susan M. Kauzlarich‡

†Nanoscale Materials Center, National Institute for Materials Science, 1-1 Namiki, Tsukuba, Ibaraki 305-0044, Japan

‡Department of Chemistry, University of California, One Shields Avenue, Davis, California 95616, U.S.A.

* To whom correspondence should be addressed. E-mail: OZAWA.Tadashi@nims.go.jp. Phone: +81-29-860-4722. FAX: +81-29-854-9061.

**Abstract**

Layered *d*-metal pnictide oxides are a unique class of compounds which consist of characteristic *d*-metal pnictide layers and metal oxide layers. More than 100 of these layered compounds, including the recently discovered Fe-based superconducting pnictide oxides, can be classified into nine structure types. These structure types and the chemical and physical properties of the characteristic *d*-metal pnictide layers and metal oxide layers of the layered *d*-metal pnictide oxides are reviewed and discussed. Furthermore, possible approaches to design new superconductors based on these layered *d*-metal pnictide oxides are proposed.

## 1. Introduction

The recent discovery of superconductivity in the Fe-based layered pnictide oxide has sparked immense interest in the chemistry and physics communities reminiscent of the discovery of the high-$T_c$ cuprate superconductors in the mid-1980's [1-3]. Pnictide oxides, which are also called "oxypnictides," are a unique class of compounds. Group 15 elements (N is often excluded) are called "pnictogens" or "pnicogens," and their anionic forms or compounds containing anionic pnictogens are called "pnictides" [4]. Pnictide ions are often found in compounds without oxygen in their components. However, when both pnictogen and oxygen coexist in a compound, they generally form polyatomic ions called "pnictates" such as phosphate ($PO_4^{3-}$), arsenate ($AsO_4^{3-}$), antimonate ($SbO_4^{3-}$) and bismuthate ($BiO_4^{3-}$), where pnictogens are cations due to the high electronegativity of the coexisting oxygen. In contrast to those pnictate-based compounds, "pnictide oxides" accommodate both pnictogen and oxygen as anions. Because of this unique mixed anionic environment, pnictide oxides tend to have characteristic structures that are rarely observed in simple oxides. For example, some of the pnictide oxides crystallize in the structure which consists of alternating fluorite- (or $ThCr_2Si_2$-) type *d*-metal (transition metal or group 12 metal) pnictide layers and $LaNiO_2$- (or square-planar $[CuO_2]^{2-}$-) type metal oxide layers interspersed with alkaline-earth metals [5-11].

Among pnictide oxides, layered *d*-metal pnictide oxides are particularly intriguing for the investigation of structure-property relationships because both the interlayer and the intralayer spacing of the system can be tuned by the appropriate selection of interlayer cations and intralayer pnictide ions. Such investigations of the structure-property relationships in the layered *d*-metal pnictide oxides began intensively in the 1990's to search for interesting and practical properties such as superconductivity [5-16]. As a matter of fact, the anomaly reminiscent of a CDW (charge-density wave) / SDW (spin-density wave) was discovered in one of the layered *d*-metal pnictide oxide families, $Na_2Ti_2Pn_2O$ (Pn = As, Sb), and generated much interest [12-15,17-20]. It has been more than a decade since these layered *d*-metal pnictide oxides were last extensively reviewed in the mid-1990's [21,22]. Since then, there have been significant developments in this field of chemistry including the discovery of many new compounds in several new structure types. Therefore, it would be interesting to newly review the layered *d*-metal pnictide oxides including the new members because such a review would promote additional research in this field. The review and discussion of the layered *d*-metal pnictide oxides are presented in terms of their structural classifications and the chemistry of their characteristic *d*-metal pnictide layers and metal oxide layers. Furthermore, possible approaches to the design of new superconductors based on these layered *d*-metal pnictide oxides are proposed.

## 2. Structures of layered *d*-metal pnictide oxides with fluorite-type [$M_2Pn_2$] layers

### *2.1. ZrCuSiAs-type pnictide oxides*

Among the pnictide oxides, those with a ZrCuSiAs-type (tetragonal, *P*4/*nmm* (No. 129)) structure have been under intense scrutiny since the recent discovery of superconductivity in LaFeOP [1]. These equiatomic quaternary pnictide oxides LnMPnO (Ln = Y, lanthanide, actinide; M = *d*-metal; Pn = pnictogen) with the ZrCuSiAs-type structure exceed 70 (table 1); thus, this is the largest family of layered *d*-metal pnictide oxides [2,23-32]. The crystal structure of ZrCuSiAs-type pnictide oxides is shown in figure 1a. This structure consists of alternating fluorite-type [$M_2Pn_2$] layers and anti-fluorite- (or $Pb_2O_2$-) type [$Ln_2O_2$] layers like that of the analogous oxychalcogenides (also called "oxide chalcogenides") such as LaAgOS [33]. The fluorite-type [$M_2Pn_2$] layer consists of square nets of M capped with Pn alternately above and below the net centers. In this layer, M is tetrahedrally coordinated by four Pn, and Pn is coordinated by four M to form square-pyramids. The anti-fluorite-type [$Ln_2O_2$] layer also has the same configuration of atoms, but it is in the reverse manner: O is tetrahedrally coordinated by four Ln, and Ln is coordinated by four O to form square-pyramids. These fluorite-type [$M_2Pn_2$] layers and anti-fluorite-type [$Ln_2O_2$] layers are also the characteristic structural features of many other types of layered *d*-metal pnictide oxides described in the following sections. In addition, the fluorite-type [$M_2Pn_2$] layers are also found in diverse non-oxide compounds, and one of their simplest forms, $ThCr_2Si_2$-type compounds, has extensively been studied both experimentally [34-38] and theoretically [39-41]. On the other hand, the anti-fluorite-type [$Ln_2O_2$] layers are often found in oxychalcogenides and oxyhalides (also called "oxide halides"), and one of their simplest forms, PbClF-type compounds, also has a large number of members [42,43]. The numerous variations of elemental combinations in this type of pnictide oxides make them suitable for investigating their structure-property relations. In general, the oxidation states of the components in this type of compounds are either ($Ln^{3+}$)($M^{2+}$)($Pn^{3-}$)($O^{2-}$) or ($Ln^{4+}$)($M^{+}$)($Pn^{3-}$)($O^{2-}$).

## 2.2. *$Th_2Ni_{3-x}P_3O$-type pnictide oxides*

The $Th_2Ni_{3-x}P_3O$-type structure crystallizes in a tetragonal cell (*P*4/*nmm* (No. 129)) with stacked fluorite-type [$M_2Pn_2$] layers and anti-fluorite-type [$Ln_2O_2$] layers like that of the ZrCuSiAs-type. However, the stacking sequence of these layers in the $Th_2Ni_{3-x}P_3O$-type pnictide oxides is –[$M_2Pn_2$]–[$M_2Pn_2$]–[$M_2Pn_2$]–[$Ln_2O_2$]– (figure 1b). In addition, the adjacent [$M_2Pn_2$] layers are linked by Pn–Pn bonds. Furthermore, a layer of Ln is captured between these linked [$M_2Pn_2$] layers. $Th_2Ni_{2.45}P_3O$ is the only pnictide oxide in this structure type reported to date (table 2) [28]. Its interlayer P–P bond distance is 2.34 Å, which is slightly longer than that in elemental phosphorous (~2.2 Å). The oxidation states of its components have been suggested to be $(Th^{4+})_2(Ni^{1+})_2(Ni^0)(P\text{–}P^{5-})(P^{3-})(O^{2-})$ where $Ni^0$ is in the middle [$M_2Pn_2$] layer among three consecutive [$M_2Pn_2$] layers with the smallest Ni occupancy. In addition, the formal charge of 5- has been assigned to the bonded P–P pairs [28].

## 2.3. *$La_3Cu_4P_4O_2$-type pnictide oxides*

The pnictide oxides of this structure type crystallize in a tetragonal cell (*I*4/*mmm* (No. 139)) with stacked fluorite-type [$M_2Pn_2$] layers and anti-fluorite-type [$Ln_2O_2$] layers in the sequence of –[$M_2Pn_2$]–[$M_2Pn_2$]–[$Ln_2O_2$]– (figure 1c). After each –[$M_2Pn_2$]–[$M_2Pn_2$]–[$Ln_2O_2$]– stacking, these layers are shifted by +1/2 along the two in-layer directions. Thus, the lattice parameter $c$ along the interlayer direction is twice as large as the thickness of the –[$M_2Pn_2$]–[$M_2Pn_2$]–[$Ln_2O_2$]– layer stack. The adjacent [$M_2Pn_2$] layers are linked by Pn–Pn bonds like those in the $Th_2Ni_{3-x}P_3O$-type pnictide oxides. There are five Ln variations with M = $Cu^+$ reported for this type of pnictide oxides $Ln_3Cu_4P_4O_{2-x}$ (Ln = La, Ce, Pr, Nd and Sm) as in table 3 [16,44]. For $Ln_3Cu_4P_4O_{2-x}$ (Ln = Pr, Sm), 25% oxygen deficiency has been inferred from the X-ray diffraction experiment [16,44]. In addition, the relatively large displacement parameter ($B$ = 3.3(9) Å$^2$) for the oxygen position of $La_3Cu_4P_4O_2$ suggests that an oxygen deficiency might also exist in $La_3Cu_4P_4O_2$ [16,44]. For Ln = Ce and Nd phases, only the lattice parameters have been reported; thus, their oxygen occupation amount is not certain. The P–P bond distance in $Pr_3Cu_4P_4O_{2-x}$ is 2.228(4) Å [44]. This P–P bond distance is within the range of typical two-electron bond distances found in the various forms of elemental P [44,45]. By considering this P–P bonding, the oxidation states of the components in $Ln_3Cu_4P_4O_{2-x}$ (Ln = Pr, Sm) have been rationalized as $(Ln^{3+})_3(Cu^{1+})_4(P{-}P^{4-})(P^{3-})_2(O^{2-})_{1.5}$, and another form of the empirical formula $Ln_6Cu_8P_8O_3$ has also been proposed [44].

## 2.4. *$U_2Cu_2As_3O$-type pnictide oxides*

$U_2Cu_2As_3O$-type pnictide oxides crystallize in a tetragonal cell (*P*4/*nmm* (No. 129)) as in figure 1d. In addition to fluorite-type [$M_2Pn_2$] layers and anti-fluorite-type [$Ln_2O_2$] layers found in the previously discussed pnictide oxide types, this type of structure contains covalently bonded square-planar [$Pn_4$] layers. This type of [$Pn_4$] layer has also been found in other pnictides like $UCuAs_2$ [46], $U_2Cu_4As_5$ [47] and $AMPn_2$ (A= Sr, Ba; M = Mn, Zn; Pn = Sb, Bi) [48,49]. However, it has only been found in the $U_2Cu_2As_3O$-type among the *d*-metal pnictide oxides. The sequence of the stacked layers in the $U_2Cu_2As_3O$-type pnictide oxides is –[$M_2Pn_2$]–[$Ln_2O_2$]–[$M_2Pn_2$]–[$Pn_4$]–. In addition, a layer of Ln is interspersed between [$M_2Pn_2$] and [$Pn_4$] layers. $U_2Cu_2As_3O$ is the only pnictide oxide in this structure type reported to date (table 2) [50]. Its As–As distance in the [$Pn_4$] layer is 2.77 Å, which is within the typical range of As–As distances (2.5 - 3.1 Å) in elemental As, and it is close to that in $UCuAs_2$ (2.79 Å) [46]. The oxidation states of the components in $U_2Cu_2As_3O$ have been proposed to be $(U^{4+})_2(Cu^{+})_2(As^{3-})_2(As^{2-})(O^{2-})$ where a charge of 2- has been assigned to the bonded As in the [$Pn_4$] layers [50]. However, the charge balance by the conduction electrons and valence band holes has also been suggested [50]. Furthermore, $U_4Cu_4P_7$, whose structure was originally reported in 1987, was later believed to be in fact a pnictide oxide $U_2Cu_2P_3O$, whose structure is closely related to this $U_2Cu_2As_3O$-type [27,28,50,51].

### *2.5. $Sr_2Mn_3As_2O_2$-type pnictide oxides*

This is a unique type of layered *d*-metal pnictide oxides containing fluorite-type [$M_2Pn_2$] layers but no anti-fluorite-type [$Ln_2O_2$] layer. The structure of this type of pnictide oxides consists of alternating fluorite-type [$M_2Pn_2$] layer and $LaNiO_2$-type square-planar [$M'O_2$] (M' = M or other *d*-metal) layer interspersed with A (alkaline-earth metal), and they crystallize in a tetragonal cell (*I*4/*mmm* (No. 139)) as in figure 1e. After each –[$M_2Pn_2$]–[M'$O_2$]– sequence, these layers are shifted by +1/2 along the two in-layer directions. This structure can also be viewed as if the $ThCr_2Si_2$-type and $LaNiO_2$-type structures are mutually intercalated. Ten pnictide oxides of this structure type have been reported to date (table 4) [5-11]. Most of the pnictide oxides in this structure type contain either $Mn^{2+}$ or $Zn^{2+}$ in both [$M_2Pn_2$] and [M'$O_2$] layers. However, ordered compounds $A_2(M_2As_2)(M'O_2)$ (A = Sr, Ba) in which $Zn^{2+}$ is only in [$M_2Pn_2$] layers and $Mn^{2+}$ is only in [M'$O_2$] layers have also been reported [10,11]. The oxidation states of the components in these compounds can be rationalized as $(A^{2+})_2(M^{2+})_2(M'^{2+})(Pn^{3-})_2(O^{2-})_2$.

## 3. Structures of layered *d*-metal pnictide oxides without fluorite-type [$M_2Pn_2$] layers

### *3.1. NdZnPO-type pnictide oxides*

This type of pnictide oxides is another polymorph of the equiatomic quaternary pnictide oxides LnMPnO [29,30]. The crystal structure of these pnictide oxides consists of alternating [$M_2Pn_2$] and [$Ln_2O_2$] layers like that of the ZrCuSiAs-type pnictide oxides (figure 2a). However, the structures of the [$M_2Pn_2$] and [$Ln_2O_2$] layers and the unit cell of the NdZnPO-type pnictide oxides are different from those of the ZrCuSiAs-type pnictide oxides. The NdZnPO-type pnictide oxides crystallize in a trigonal cell (*R*-3*m* (No. 166)). The [$Ln_2O_2$] layers of the NdZnPO-type pnictide oxides are identical to those in the $Ce_2O_2S$-type pnictide oxides [52]. It consists of a network of O-centered tetrahedra coordinated by four Ln, and each Ln is coordinated by three O at the corners of a trigonal base and one more O alternately down or up toward the trigonal base center. On the other hand, the [$M_2Pn_2$] layer of the NdZnPO-type pnictide oxides is anti-$Ce_2O_2S$-type. This type of the layer consists of a network of M-centered tetrahedra coordinated by four Pn, and each Pn is coordinated by three M at the corners of a trigonal base and one more M alternately down or up toward the trigonal base center. More than ten pnictide oxides in this structure type have been reported (table 5). Some of them, such as CeZnPO and PrZnPO, can have both ZrCuSiAs- and NdZnPO-type structures. In such a case, ZrCuSiAs- and NdZnPO-types are the low- and high-temperature phases, respectively [30]. In general, the oxidation states of the components in these compounds are ($Ln^{3+}$)($M^{2+}$)($Pn^{3-}$)($O^{2-}$).

*3.2. $Na_2Ti_2Sb_2O$-type pnictide oxides*

This type of pnictide oxides crystallize in a tetragonal (*I*4/*mmm* (No. 139)) structure with [$M_2Pn_2O$] layers interspersed with double layers of A' (alkali metal) as in figure 2b [12-15, 53]. The connectivity of atoms in this type of [$M_2Pn_2O$] layer is quite unique. It has a $M_2O$ square net that is anti-structural to the $LaNiO_2$-type [$CuO_2$] net of the cuprate superconductors. M in this layer is also coordinated by four Pn located above and below the center of the $M_2O$ squares to form M-centered $MPn_4O_2$ octahedra. This structure can also be viewed as the ordered anti-$K_2NiF_4$-type where A' and M are at the ordered F site, Pn is at the K site, and O is at the Ni site of the $K_2NiF_4$-type structure. $Na_2Ti_2As_2O$ and $Na_2Ti_2Sb_2O$ have been found to crystallize in this structure type (table 6) [53]. The oxidation states of the components in these compounds can be rationalized as $(A'^{+})_2(M^{3+})_2(Pn^{3-})_2(O^{2-})$.

*3.3. $Ba_2Mn_2Sb_2O$-type pnictide oxides*

This is the only family among the layered *d*-metal pnictide oxides crystallizing in a hexagonal structure ($P6_3/mmc$ (No. 194)) as in figure 2c [54]. In this structure, M is tetrahedrally coordinated by three Pn and one O. These distorted tetrahedra are corner-shared at the Pn sites to form a two-dimensional network. The tetrahedra are also corner-shared at the O sites to form a double tetrahedra layer [$M_2Pn_2O$]. Furthermore, half of A is enclosed within these [$M_2Pn_2O$] layers as [$AM_2Pn_2O$]. The overall structure of this type of pnictide oxides consists of stacked [$AM_2Pn_2O$] layers interspersed with A. $Ba_2Mn_2Sb_2O$ and $Ba_2Mn_2Bi_2O$ have been found to crystallize in this structure type (table 6) [54]. The oxidation states of the components in these compounds can be rationalized as $(A^{2+})_2(M^{2+})_2(Pn^{3-})_2(O^{2-})$.

### *3.4. $Ba_2Mn_2As_2O$-type pnictide oxides*

This is another unique type of the layered *d*-metal pnictide oxides that crystallize in a monoclinic structure ($I2/m$ (No. 12)) as in figure 2d [55]. This structure consists of corner- and edge-shared distorted tetrahedra. These tetrahedra have M at the center, which is coordinated by three Pn and one O, like those in the $Ba_2Mn_2Sb_2O$-type [54]. However, the connectivity of these tetrahedra in the $Ba_2Mn_2As_2O$-type pnictide oxide is different from that in the $Ba_2Mn_2Sb_2O$-type. In the $Ba_2Mn_2As_2O$-type, these tetrahedra have a corner-shared linkage at the O sites, and an edge-shared linkage at the Pn$\cdots$Pn edges. Only $Ba_2Mn_2As_2O$ has been found to crystallize in this structure type (table 6) [55]. The oxidation states of the components in this compound can be rationalized as $(A^{2+})_2(M^{2+})_2(Pn^{3-})_2(O^{2-})$.

## 4. Chemical and physical properties of characteristic *d*-metal pnictide layers and metal oxide layers

In the preceding sections, unique structural features of the layered *d*-metal pnictide oxides were reviewed and described. The atoms within these layers are bonded providing a two-dimensional nature whereas the interlayer interactions are ionic. The atomic connectivity in these characteristic layers and the coexistence of different types of layers in many of the layered *d*-metal pnictide oxides is quite intriguing. Furthermore, the chemical and physical properties of these compounds are primarily governed by their characteristic *d*-metal pnictide layers and metal oxide layers. In this section, the chemical and physical properties of these characteristic *d*-metal pnictide layers and metal oxide layers in the layered *d*-metal pnictide oxides are reviewed.

### *4.1. Fluorite-type [$M_2Pn_2$] layers*

The layered *d*-metal pnictide oxides with the fluorite-type [$M_2P_2$] layers can be further categorized into two groups depending on whether or not they contain a Pn–Pn bond. The first group consists of the ZrCuSiAs- (figure 1a) and $Sr_2Mn_3As_2O_2$- (figure 1e) type pnictide oxides, which do not contain any Pn–Pn bond. For the ZrCuSiAs-type pnictide oxides, the oxidation state of M is 2+ ($Mn^{2+}$, $Fe^{2+}$, $Co^{2+}$, $Ni^{2+}$, $Zn^{2+}$, $Ru^{2+}$, $Cd^{2+}$) if the oxidation state of Ln is 3+, and the oxidation state of M is 1+ ($Cu^{+}$) if the oxidation state of Ln is 4+. For the $Sr_2Mn_3As_2O_2$-type pnictide oxides, the oxidation state of M is 2+ ($Mn^{2+}$, $Zn^{2+}$). The other group consists of $Th_2Ni_{3-x}P_3O$- (figure 1b), $La_3Cu_4P_4O_2$- (figure 1c) and $U_2Cu_2As_3O$- (figure 1d) types which contain Pn–Pn bonds in either the inter-[$M_2Pn_2$] layer linkage or the square-planar [$Pn_4$] layers. For this second group, the oxidation state of M is 1+ ($Ni^{+}$, $Cu^{+}$). In the case of ZrCuSiAs-type oxysulfides such as LaAgOS, $Ag^{+}$ has been observed in the M site of the fluorite-type [$M_2S_2$] layers [33]. Thus, it might be possible to accommodate $Ag^+$ in the fluorite-type [$M_2Pn_2$] layers of new pnictide oxides. The pnictogen anions $P^{3-}$, $As^{3-}$ and $Sb^{3-}$ are often observed in the layered *d*-metal pnictide oxides containing the fluorite-type [$M_2Pn_2$] layers. However, $Bi^{3-}$ has only been found in the $Sr_2Mn_3As_2O_2$-type $Sr_2Mn_3Bi_2O_2$ among the pnictide oxides with the fluorite-type [$M_2Pn_2$] layers [5]. The average in-layer lattice parameter of the previously reported $Sr_2Mn_3As_2O_2$-type pnictide oxides is 4.22 Å, which is larger than those of any other types of tetragonal layered *d*-metal pnictide oxides (expect for the ZrCuSiAs-type LaMnSbO and LaZnSbO whose $a$ = 4.242(1) and 4.2262(2) Å, respectively) [5-11,31]. The coexistence of the relatively large $LaNiO_2$-type [M'O] layer with the fluorite-type [$M_2Pn_2$] layer is likely to be the reason for the accommodation of $Bi^{3-}$ in the fluorite-type [$M_2Pn_2$] layer of the $Sr_2Mn_3As_2O_2$-type pnictide oxide. In general, $Bi^{3-}$ might be too large to be accommodated in the fluorite-type [$M_2Pn_2$] layers of other types of pnictide oxides, and the less electronegative nature of Bi with respect to that of other pnictogens might also be preventing the accommodation of $Bi^{3-}$ in the layers of the majority of pnictide oxides.

The magnetic properties of the $Sr_2Mn_3As_2O_2$-type pnictide oxides with the fluorite-type [$Mn_2Pn_2$] layers have been studied, and antiferromagnetic interactions of $Mn^{2+}$ were observed [6-8]. In addition, the ZrCuSiAs-type pnictide oxides with fluorite-type [$Zn_2P_2$] layers have been reported to be black in color, and their metallic nature has been attributed to their significant M–M interactions [24,25,29,32]. Furthermore, superconductivity in the ZrCuSiAs-type LaMPO (M = Fe, Ni) and $LnFeAsO_{1-x}F_x$ with fluorite-type [$M_2Pn_2$] layers has recently been discovered [1,2, 26]. Also, the $La_3Cu_4P_4O_2$-type pnictide oxides $Ln_3Cu_4P_4O_{2-x}$ (Ln = La, Ce, Pr, Nd,

Sm), which contain fluorite-type [$Cu_2P_2$] layers, are reported to be metallic or have metallic luster [16,44].

*4.2. Anti-fluorite-type [$Ln_2O_2$] layers*

Anti-fluorite-type [$Ln_2O_2$] layers have been found in the ZrCuSiAs- (figure 1a), $Th_2Ni_{3-x}P_3O$- (figure 1b), $La_3Cu_4P_4O_2$- (figure 1c) and $U_2Cu_2As_3O$- (figure 1d) type pnictide oxides. Ln can be Y, lanthanide or actinide whose oxidation states are either 3+ or 4+. However, $Eu^{3+}$ has not been found in this type of layers among the layered *d*-metal pnictide oxides even though it can be observed in the anti-fluorite-type [$Ln_2O_2$] layers in oxides like $Eu_2CuO_4$ [56]. In addition, lanthanide 3+ ions with sizes larger or smaller than that of $Eu^{3+}$ have been observed in the anti-fluorite-type [$Ln_2O_2$] layers of the layered *d*-metal pnictide oxides. These results indicate that the anionic environment of the pnictide oxide is not sufficiently electronegative to oxidize Eu to 3+ because Eu is also relatively stable as 2+. Similarly, $Bi^{3+}$ has not been observed in the anti-fluorite-type [$Ln_2O_2$] layer of the layered *d*-metal pnictide oxides even though it has been observed in this type of layers among oxides like Aurivillius phases [57,58] and oxychalcogenides like BiCuOCh (Ch = S, Se) [59]. Furthermore, the coexistence of $Pn^{3+}$ and $Pn^{3-}$ in pnictide oxides has not been reported to the best of our knowledge. These trends also indicate that the anionic environment of the pnictide oxide is not sufficiently electronegative to oxidize part of the Pn to $Pn^{3+}$. We expect that all the Pn might be oxidized to $Pn^{3+}$ to form pnictate ions if a highly oxidative synthetic approach is employed. In addition, the smallest lanthanide in the anti-fluorite-type [$Ln_2O_2$] layers among the pnictide oxides is Dy in the ZrCuSiAs-type pnictide oxides [23,25,29]. From these facts, a subtle balance between the electronegativity and the size of Ln in the anti-fluorite-type [$Ln_2O_2$] layers of the pnictide oxides is inferred.

The magnetic properties of the $La_3Cu_4P_4O_2$-type pnictide oxides with anti-fluorite-type [$Ln_2O_2$] layers depend on the kind of Ln ions. The magnetic susceptibility of $La_3Cu_4P_4O_2$ is dominated by core diamagnetism and weak Pauli paramagnetism from the conduction electrons [16]. The magnetic susceptibility of $Ce_3Cu_4P_4O_2$ exhibits temperature dependent curvature, and the fit of the susceptibility to the Curie-Weiss law yields the Ce moment of 2.33 $\mu_B$/Ce and $\theta$ (Weiss temperature) of 39.2 K [16]. In addition, the magnetic susceptibility of $Nd_3Cu_4P_4O_2$ also exhibits the Curie-Weiss trend corresponding to the Nd moment of 3.68 $\mu_B$/Nb and $\theta$ of 23.0 K, suggesting the existence of antiferromagnetic interactions [16]. However, no magnetic ordering has been observed above 4.2 K for $Ce_3Cu_4P_4O_2$ and $Nd_3Cu_4P_4O_2$ [16].

*4.3. $LaNiO_2$-type [$MO_2$] layers*

The extensive studies were performed during the 1990's to search for new pnictide oxides containing $LaNiO_2$-type [$M'O_2$] layers [6-11,31,60,61]. A particular interest at that time was to search for $[CuO_2]^{2-}$-layer-containing pnictide oxides because many compounds containing $[CuO_2]^{2-}$ layers are known to exhibit high-$T_c$ superconductivity [3,11,16,61]. However, pnictide oxides with $Cu^{2+}$ might be quite difficult to prepare. The oxidation state of Cu found in the previously reported pnictide oxides is always 1+. The oxidation potential of Cu is much lower than those of other *d*-metals like Mn and Zn, which are commonly observed in pnictide oxides. This might indicate that the anionic environment of the pnictide oxides is in general not sufficiently electronegative to oxidize Cu to 2+. Even in the $LaNiO_2$-type [$M'O_2$] layer of the $Sr_2Mn_3As_2O_2$-type oxysulfides $Sr_2[M'_{1-x}Cu^{2+}_xO_2][Cu^{1+}_2S_2]$ (M' = Sc, Cr, Mn, Fe, Co, Ni, Zn), which are expected to have more electronegative anionic environment than the pnictide oxides, the maximum $Cu^{2+}$ content for the single-phase sample was $x < 1$ [62-66]. This also indicates that the accommodation of $Cu^{2+}$ in the weakly electronegative anionic environment of suboxides such as pnictide oxides and oxychalcogenides is difficult in contrast to its accommodation in the more electronegative anionic environment of oxides.

$LaNiO_2$-type $[M'O_2]$ layers have only been observed in the $Sr_2Mn_3As_2O_2$-type pnictide oxides with M = $Mn^{2+}$ or $Zn^{2+}$ among all the pnictide oxides, and they are all semiconducting (figure 1e and table 4) [5-11,21,22]. The *d*-metal site (M or M') selectivity has been investigated for the $Sr_2Mn_3As_2O$-type pnictide oxides $AM_{3-x}M'_xAs_2O_2$ (A = Sr, Ba), which consist of alternating fluorite-type $[M_2Pn_2]$ layers and $LaNiO_2$-type $[M'O_2]$ layers (figure 1e). For M = $Mn^{2+}$ and $Zn^{2+}$, both end members $A_2Mn_3As_2O_2$ and $A_2Zn_3As_2O_2$ exist [5,9]. However, at $x = 1$, the structure orders where $Zn^{2+}$ occupies the tetrahedral site of the fluorite-type $[M_2Pn_2]$ layers and $Mn^{2+}$ occupies the square-planar site of the $LaNiO_2$-type $[M'O_2]$ layers as $A_2(Zn_2As_2)(MnO_2)$ [10,11]. The site order in this type of pnictide oxides is considered to be mainly due to the tetrahedral site preference of $Zn^{2+}$, whose occupancy in a square-planar site is rarely observed [11]. It might be possible to design new ordered *d*-metal layered compounds utilizing this kind of site preference of *d*-metals and a $Sr_2Mn_3As_2O_2$-type pnictide oxide as a template. In addition, spin-glass-like magnetic interactions were observed for this ordered $Sr_2Mn_3As_2O$-type pnictide oxides with the $LaNiO_2$-type $[MnO_2]$ layer, and generated interest from the structure-property viewpoint [10,11,60,67,68].

*4.4. $Ce_2O_2S$-type [$Ln_2O_2$] layers*

In contrast to the anti-fluorite-type [$Ln_2O_2$] layers, $Ce_2O_2S$-type [$Ln_2O_2$] layers can accommodate a small lanthanide such as Tm. However, a large lanthanide La has not been found in this type of layers among the layered *d*-metal pnictide oxides. These results imply that the $Ce_2O_2S$-type [$Ln_2O_2$] layer preferentially accommodates smaller lanthanides than the anti-fluorite-type [$Ln_2O_2$] layer. The $Ce_2O_2S$-type [$Ln_2O_2$] layers are only found in the NdZnPO- type structure (figure 2a and table 5) among the layered *d*-metal pnictide oxides. However, they have also been found in the $Ce_2O_2S$-type pnictide oxides $Th_2(N,O)_2Pn$ (Pn = P, As) with no *d*-metal [52].

*4.5. Anti-$Ce_2O_2S$-type [$M_2Pn_2$] layers*

Anti-$Ce_2O_2S$-type [$M_2Pn_2$] layers have only been found in the NdZnPO-type (figure 2a and table 5) among the pnictide oxides [29,30] even though this type of layers have been observed in many intermetallic compounds [69]. In addition, only M = $Zn^{2+}$ has been found in the anti-$Ce_2O_2S$-type [$M_2Pn_2$] layers of the NdZnPO-type pnictide oxides. This implies that the anti-$Ce_2O_2S$-type [$M_2Pn_2$] layers preferentially accommodate small M. Similarly, only Pn = $P^{3-}$ has been observed in the anti-$Ce_2O_2S$-type [$M_2Pn_2$] layers of the NdZnPO-type pnictide oxides. The NdZnPO-type pnictide oxides are transparent implying their insulating nature [29]. From this result, the M–M interaction in the NdZnPO-type pnictide oxides is expected to be much less than that in the ZrCuSiAs-type pnictide oxides in which their metallic property has been attributed to the significant M–M bonding [29].

*4.6. $Na_2Ti_2Sb_2O$-type [$M_2Pn_2O$] layers*

Among the layered *d*-metal pnictide oxides, this type of layer has only been observed in the $Na_2Ti_2Pn_2O$-type pnictide oxides [53] even though quite similar layers have also been observed in the anti-$K_2NiF_4$-type pnictide oxides [70-79] and the anti-Ruddlesden-Popper-type $(Ba_{0.62}Sr_{0.38})_{10}N_2OBi_4$ pnictide oxide [80] with no *d*-metal. Among the pnictogens, only As and Sb have been observed in the $Na_2Ti_2Pn_2O$-type [$M_2Pn_2O$] layer of the layered *d*-metal pnictide oxides (figure 2b and table 6) [53]. Furthermore, the only *d*-metal observed in the $Na_2Ti_2Sb_2O$-type [$M_2Pn_2O$] layer is $Ti^{3+}$ in $Na_2Ti_2Pn_2O$ (Pn = As, Sb) [53]. However, the layers with the same arrangement of *d*-metal, oxygen and chalcogen, which is in the position of Pn, have also been observed in oxychalcogenides such as $La_2Fe_2O_3Ch_2$ (Ch = S, Se) [81].

$Na_2Ti_2As_2O$ and $Na_2Ti_2Sb_2O$ exhibit transitions reminiscent of CDW/SDW, in both temperature-dependent magnetic susceptibility and electrical resistivity around 330 and 120 K, respectively, and this discovery of an exotic low-dimensional property generated much interest among chemists and physicists [12-15, 17-20]. A similar anomaly has also been reported in the previously mentioned suboxides $La_2Fe_2O_3Ch_2$ (Ch = S, Se) with $Na_2Ti_2Sb_2O$-type [$Fe_2Ch_2O$] layers. Thus, it is likely that there is a strong correlation between $Na_2Ti_2Sb_2O$-type [$M_2Su_2O$] (Su = Pn, Ch) layers and the CDW/SDW-like anomaly. In addition, a large single crystal growth method, which utilizes a binary alkali metal pnictide as flux, has been developed for $Na_2Ti_2Sb_2O$ [15].

*4.7. $Ba_2Mn_2Sb_2O$- and $Ba_2Mn_2As_2O$-type [$M_2Pn_2O$] layers*

$Ba_2Mn_2Sb_2O$-type [$M_2Pn_2O$] layers have been observed in $Ba_2Mn_2Pn_2O$ (Pn = Sb, Bi) (figure 2c) [54] whereas $Ba_2Mn_2As_2O$-type [$M_2Pn_2O$] layers have only been observed in $Ba_2Mn_2As_2O$ (figure 2d) [55] as in table 6. Considering the fact that these two types of layers have the same composition of M, Pn and O, it is rational to expect that the $Ba_2Mn_2Sb_2O$-type [$M_2Pn_2O$] layer preferentially accommodates larger Pn than the $Ba_2Mn_2As_2O$-type [$M_2Pn_2O$] layer. It is interesting to know that only the $Ba_2Mn_2Sb_2O$-type [$M_2Pn_2O$] layer and the fluorite-type [$M_2Pn_2$] layer have been reported to accommodate $Bi^{3-}$ among all types of layers in the layered *d*-metal pnictide oxides. In addition, only $Mn^{2+}$ has been found in the M sites of $Ba_2Mn_2Sb_2O$- and $Ba_2Mn_2As_2O$-type [$M_2Pn_2O$] layers. Considering the variety of tetrahedrally coordinated M in other types of layers in the *d*-metal pnictide oxides, many other pnictide oxides with the $Ba_2Mn_2Sb_2O$- and $Ba_2Mn_2As_2O$-type [$M_2Pn_2O$] layers might exist for M other than $Mn^{2+}$.

## 5. Layered *d*-metal pnictide oxides as candidates for new superconductors

The recent discovery of the Fe-based ZrCuSiAs-type pnictide oxide superconductors has sparked the renewed interest in layered *d*-metal pnictide oxides [1,2]. In this system, superconductivity originates from the fluorite-type [$Fe_2Pn_2$] layers. It is generally believed that magnetic elements such as Fe tend to annihilate superconductivity. Some of the Fe-containing superconductors, such as $LnFe_4P_{12}$ (Ln = Y and La) [82,83], have previously been known, but their $T_c$ (superconducting critical temperature) has been much lower than that of the ZrCuSiAs-type pnictide oxide $LaFeAsO_{1-x}F_x$ at 26 K [2]. The superconductivity in the layered *d*-metal pnictide oxides and high-$T_c$ cuprates is quite similar in the following aspects. First, the structures of these superconductors consist of two-dimensional superconducting layers and blocking layers, and superconductivity is induced by either electron or hole carrier doping through aliovalent substitution or adjusting the anion content. For example, carrier doping by the partial replacement of $O^{2-}$ with $F^-$ (electron doping) or $La^{3+}$ with $Sr^{2+}$ (hole doping) in the anti-fluorite-type [$La_2O_2$] blocking layer was necessary in order to induce the superconductivity in the fluorite-type [$Fe_2As_2$] layer of LaFeAsO [1,84]. Second, $T_c$ of these superconductors can be raised by the application of pressure. For example, $T_c$ of $LaFeAsO_{1-x}F_x$ can be raised to as high as 43 K under pressure [85]. Third, the replacement of a constituting ion with a smaller one (the application of chemical pressure) tends to raise $T_c$ of these superconductors. For example, $T_c$ of $LaFeAsO_{1-x}F_x$ has been raised up to 55 K by the replacement of La with a smaller lanthanide such as Sm in the anti-fluorite-type [$Ln_2O_2$] layer [86]. Finally, a variety of compounds containing the fundamental superconducting layers, which are the fluorite-type [$M_2Pn_2$] (M = Fe, Ni) layers for pnictide oxides and the $LaNiO_2$-type [$CuO_2$] layers for cuprates, are expected to exhibit superconductivity. The fluorite-type [$Fe_2As_2$] layer as the crucial structural factor for superconductivity in these pnictide oxides has been manifested by the recent discovery of superconductivity at 30 K in $Ba_{1-x}K_xFe_2As_2$, which has one of the simplest fluorite-type [$Fe_2As_2$] layer-containing structures [87]. As described in this review, many of the layered pnictide oxides other than those with the ZrCuSiAs-type structure also contain fluorite-type [$M_2Pn_2$] layers; thus, they are possible candidates for new superconductors.

Among these fluorite-type-[$M_2Pn_2$] layer-containing pnictide oxides, the $Sr_2Mn_3As_2O_2$-type pnictide oxides (figure 1e) are excellent candidates for new superconductors for the following reasons. First, the oxidation state of M is 2+ in the fluorite-type [$M_2Pn_2$] layer of the $Sr_2Mn_3As_2O_2$-type pnictide oxides; thus, the incorporation of $Fe^{2+}$ and the formation of a superconducting [$Fe_2Pn_2$] layer appear to be feasible. Second, charge carriers might be introduced by chemical doping into either the interlayer A site or the $LaNiO_2$-type [$M'O_2$] layer sites. Thus, it would be interesting to synthesize new $Sr_2Mn_3As_2O_2$-type pnictide oxides such as $A_2Fe_3Pn_2O_2$ (A = Sr, Ba) and investigate the effects of electron and hole doping. In addition, it might also be necessary to replace the M' site in the $LaNiO_2$-type [$M'O_2$]-layer of such a pnictide oxide with a *d*-metal other than Fe in order to induce superconductivity. Such site-selective doping is possible because the M site in the fluorite-type [$M_2Pn_2$] layer and the M' site in the $LaNi_2O_2$-type [$M'O_2$] layer of $Sr_2Mn_3As_2O_2$-type pnictide oxides have tetrahedral and square-planar coordination, respectively, and the site-preference properties of *d*-metals can be utilized as described in section 4.3.

Other possible candidates for new superconductors are $Na_2Ti_2Sb_2O$-type pnictide oxides (figure 1b). The anomaly reminiscent of CDW/SDW has been observed in the electrical resistivity and magnetic susceptibilities of $Na_2Ti_2As_2O$ and $Na_2Ti_2Sb_2O$ [12-14]. The electron-phonon and/or electron-spin interactions in CDW/SDW materials are similar to those in superconductors [88-90]. The superconducting state is often stabilized or induced by annihilating the CDW/SDW states [89,90]. In fact, the annihilation of the SDW state in LaFeAsO by carrier doping was crucial in order to induce superconductivity [91,92]. From this viewpoint, it is rational to expect that the annihilation of the CDW/SDW state by carrier doping or the application of pressure might induce superconductivity in $Na_2Ti_2Pn_2O$.

## 6. Conclusion

The layered *d*-metal pnictide oxides are a unique class of compounds which are composed of distinctive *d*-metal oxide layers and metal pnictide layers. Among the 90 naturally existing elements, only about 30 elements have been observed in less than 10 structure types for these compounds. However, over 100 of these exotic compounds have been reported with various compositions of these particular elements. The chemistry of the layered *d*-metal pnictide oxides is still relatively unexplored. Thus, more of these unique compounds are likely to be discovered in the future. In addition, the physical properties of many of these pnictide oxides have not been well characterized. The recent discovery of superconductivity in the previously reported ZrCuSiAs-type pnictide oxide LaFePO exemplifies this fact [1,24]. This suggests that the layered *d*-metal pnictide oxides are a possible treasure trove of unexpected practical materials.

## Acknowledgements

This work was partially funded by NSF (DMR). All the crystal structures were drawn with **Balls and Sticks**, a free crystal structure visualization software [93].

**Table 1. Unit Cell Parameters of ZrCuSiAs-type Pnictide Oxides**

| Compound | *a* (Å) | *c* (Å) | Ref |
|---|---|---|---|
| LaMnPO | 4.054(1) | 8.834(4) | [23] |
| CeMnPO | 4.020(1) | 8.742(3) | [23] |
| PrMnPO | 4.006(1) | 8.707(2) | [23] |
| NdMnPO | 3.989(1) | 8.674(1) | [23] |
| SmMnPO | 3.960(1) | 8.590(3) | [23] |
| GdMnPO | 3.933(1) | 8.510(1) | [23] |
| TbMnPO | 3.920(1) | 8.485(4) | [23] |
| DyMnPO | 3.904(1) | 8.469(4) | [23] |
| YMnAsO | 3.957(1) | 8.750(6) | [23] |
| LaMnAsO | 4.124(1) | 9.030(5) | [23] |
| CeMnAsO | 4.086(1) | 8.956(2) | [23] |
| PrMnAsO | 4.067(1) | 8.919(3) | [23] |
| NdMnAsO | 4.049(2) | 8.893(1) | [23] |
| SmMnAsO | 4.020(1) | 8.829(3) | [23] |
| GdMnAsO | 3.989(1) | 8.805(3) | [23] |
| TbMnAsO | 3.978(1) | 8.743(4) | [23] |
| DyMnAsO | 3.959(1) | 8.727(4) | [23] |
| UMnAsO | 3.869(1) | 8.525(2) | [23] |
| LaMnSbO | 4.242(1) | 9.557(2) | [23] |
| CeMnSbO | 4.218(1) | 9.517(2) | [23] |
| PrMnSbO | 4.187(1) | 9.460(1) | [23] |
| NdMnSbO | 4.165(1) | 9.462(2) | [23] |
| SmMnSbO | 4.135(1) | 9.418(2) | [23] |
| GdMnSbO | 4.090(1) | 9.410(1) | [23] |
| LaFePO | 3.9570(9) | 8.507(4) | [24] |
| CeFePO | 3.919(1) | 8.327(3) | [24] |
| PrFePO | 3.9113(6) | 8.345(2) | [24] |
| NdFePO | 3.8995(5) | 8.302(3) | [24] |
| SmFePO | 3.878(1) | 8.205(1) | [24] |
| GdFePO | 3.861(3) | 8.123(7) | [24] |
| LaFeAsO | 4.038(1) | 8.753(6) | [25] |
| $LaFeAsO_{0.95}F_{0.05}$ | 4.0320(1) | 8.7263(3) | [2] |
| CeFeAsO | 4.000(1) | 8.655(1) | [25] |
| PrFeAsO | 3.985(1) | 8.595(3) | [25] |
| NdFeAsO | 3.965(1) | 8.575(2) | [25] |
| SmFeAsO | 3.940(1) | 8.496(3) | [25] |
| GdFeAsO | 3.915(1) | 8.435(4) | [25] |
| LaCoPO | 3.9678(9) | 8.379(3) | [24] |
| CeCoPO | 3.9213(7) | 8.219(4) | [24] |
| PrCoPO | 3.9224(8) | 8.224(2) | [24] |
| NdCoPO | 3.9084(5) | 8.172(2) | [24] |
| SmCoPO | 3.8817(7) | 8.073(2) | [24] |
| LaCoAsO | 4.054(1) | 8.472(3) | [25] |
| CeCoAsO | 4.015(1) | 8.364(2) | [25] |
| PrCoAsO | 4.005(1) | 8.344(2) | [25] |
| NdCoAsO | 3.982(1) | 8.317(4) | [25] |
| LaNiPO | 4.0461(8) | 8.100(7) | [26] |
| UCuPO | 3.793(1) | 8.233(2) | [27] |
| $ThCu_{1-x}PO$ | 3.8943(4) | 8.283(1) | [28] |
| ThCuAsO | 3.9614(5) | 8.440(1) | [28] |
| LaZnPO | 4.040(1) | 8.908(2) | [29] |
| CeZnPO (low-temperature phase) | 4.013(1) | 8.824(2) | [29] |
| PrZnPO (low-temperature phase) | 3.993(2) | 8.772(7) | [30] |
| YZnAsO | 3.943(1) | 8.843(3) | [29] |
| LaZnAsO | 4.095(1) | 9.068(3) | [29] |
| CeZnAsO | 4.069(1) | 8.995(3) | [29] |
| PrZnAsO | 4.047(1) | 8.963(1) | [29] |
| NdZnAsO | 4.030(1) | 8.949(4) | [29] |
| SmZnAsO | 4.003(1) | 8.903(2) | [29] |
| GdZnAsO | 3.976(1) | 8.894(3) | [29] |
| TbZnAsO | 3.957(1) | 8.841(2) | [29] |
| DyZnAsO | 3.947(1) | 8.838(1) | [29] |
| LaZnSbO | 4.2262(2) | 9.5377(6) | [31] |
| CeZnSbO | 4.1976(4) | 9.474(1) | [31] |
| PrZnSbO | 4.1763(4) | 9.451(1) | [31] |
| NdZnSbO | 4.1581(2) | 9.4495(5) | [31] |
| SmZnSbO | 4.1280(2) | 9.4016(6) | [31] |
| LaRuPO | 4.047(1) | 8.406(1) | [24] |
| CeRuPO | 4.026(1) | 8.256(2) | [24] |
| PrRuPO | 4.018(1) | 8.174(3) | [24] |
| NdRuPO | 4.0086(5) | 8.167(2) | [24] |
| SmRuPO | 3.9935(5) | 8.058(2) | [24] |
| GdRuPO | 3.979(1) | 7.974(2) | [24] |
| LaRuAsO | 4.119(1) | 8.488(1) | [25] |
| CeRuAsO | 4.096(1) | 8.380(3) | [25] |
| PrRuAsO | 4.085(1) | 8.337(1) | [25] |
| NdRuAsO | 4.079(1) | 8.292(2) | [25] |
| SmRuAsO | 4.050(2) | 8.191(7) | [25] |
| GdRuAsO | 4.039(1) | 8.118(6) | [25] |
| TbRuAsO | 4.027(1) | 8.078(1) | [25] |
| DyRuAsO | 4.022(2) | 8.050(3) | [25] |
| LaCdPO | 4.172(2) | 9.067(6) | [32] |
| LaCdAsO | 4.129(2) | 9.230(3) | [32] |
| CeCdAsO | 4.191(1) | 9.171(4) | [32] |
| PrCdAsO | 4.172(1) | 9.136(4) | [32] |
| NdCdAsO | 4.151(1) | 9.123(5) | [32] |
| (Nd, Sm)CdAsO | 4.147(2) | 9.125(6) | [32] |

**Table 2. Unit Cell Parameters of $Th_2Ni_{3-x}P_3O$- and $U_2Cu_2As_3O$-type Pnictide Oxides**

| Compound | *a* (Å) | *c* (Å) | Ref |
|---|---|---|---|
| $Th_2Ni_{2.45}P_3O$ | 3.9462(4) | 17.232(3) | [28] |
| $U_2Cu_2As_3O$ | 3.9111(2) | 17.916(4) | [50] |

**Table 3. Unit Cell Parameters of $La_3Cu_4P_4O_2$-type Pnictide Oxides**

| Compound | *a* (Å) | *c* (Å) | Ref |
|---|---|---|---|
| $La_3Cu_4P_4O_2$ | 4.033(1) | 26.765(8) | [16] |
| $Ce_3Cu_4P_4O_2$ | 3.985(1) | 26.573(9) | [16] |
| $Pr_3Cu_4P_4O_{2-x}$ | 3.978(1) | 26.587(3) | [44] |
| $Nd_3Cu_4P_4O_2$ | 3.964(1) | 26.551(5) | [16] |
| $Sm_3Cu_4P_4O_{2-x}$ | 3.928(1) | 26.436(3) | [44] |

**Table 4. Unit Cell Parameters of $Sr_2Mn_3As_2O_2$-type Pnictide Oxides**

| Compound | *a* (Å) | *c* (Å) | Ref |
|---|---|---|---|
| $Sr_2Mn_3As_2O_2$ | 4.16(1) | 18.84(4) | [5] |
| $Sr_2Mn_3Sb_2O_2$ | 4.262(4) | 20.11(2) | [5] |
| $Sr_2Mn_3Bi_2O_2$ | 4.28(1) | 20.55(5) | [5] |
| $Ba_2Mn_3P_2O_2$ | 4.2029(7) | 19.406(5) | [6] |
| $Ba_2Mn_3As_2O_2$ | 4.248(5) | 19.77(3) | [5] |
| $Ba_2Mn_3Sb_2O_2$ | 4.367(5) | 20.78(2) | [5] |
| $Sr_2Zn_3As_2O_2$ | 4.0954(7) | 18.918(4) | [9] |
| $Ba_2Zn_3As_2O_2$ | 4.2202(3) | 19.713(4) | [9] |
| $Sr_2MnZn_2As_2O_2$ | 4.126237(24) | 18.67091(16) | [11] |
| $Ba_2MnZn_2As_2O_2$ | 4.23369(4) | 19.4780(3) | [10] |

**Table 5. Unit Cell Parameters of NdZnPO-type Pnictide Oxides**

| Compound | *a* (Å) | *c* (Å) | Ref |
|---|---|---|---|
| YZnPO | 3.883(1) | 30.319(7) | [29] |
| CeZnPO (high temperature phase) | 4.012(4) | 31.21(3) | [30] |
| PrZnPO (high temperature phase) | 3.986(1) | 31.054(4) | [29] |
| NdZnPO | 3.977(1) | 30.975(5) | [29] |
| SmZnPO | 3.948(1) | 30.749(6) | [29] |
| GdZnPO | 3.918(1) | 30.548(6) | [29] |
| TbZnPO | 3.904(1) | 30.447(9) | [29] |
| DyZnPO | 3.891(1) | 30.324(6) | [29] |
| HoZnPO | 3.882(1) | 30.249(6) | [29] |
| ErZnPO | 3.868(1) | 30.150(6) | [29] |
| TmZnPO | 3.859(1) | 30.079(6) | [29] |

**Table 6. Unit Cell Parameters of $Na_2Ti_2Sb_2O$-, $Ba_2Mn_2Sb_2O$- and $Ba_2Mn_2As_2O$-type Pnictide Oxides**

| Compound | type | *a* (Å) | *b* (Å) | *c* (Å) | Ref |
|---|---|---|---|---|---|
| $Na_2Ti_2As_2O$ | $Na_2Ti_2Sb_2O$ | 4.070(2) | = *a* | 15.288(4) | [53] |
| $Na_2Ti_2Sb_2O$ | $Na_2Ti_2Sb_2O$ | 4.144(0) | = *a* | 16.561(1) | [53] |
| $Ba_2Mn_2Sb_2O$ | $Ba_2Mn_2Sb_2O$ | 4.71(1) | = *a* | 20.04(2) | [54] |
| $Ba_2Mn_2Bi_2O$ | $Ba_2Mn_2Sb_2O$ | 4.803(5) | = *a* | 20.097(10) | [54] |
| $Ba_2Mn_2As_2O^*$ | $Ba_2Mn_2As_2O$ | 7.493(4) | 4.196(1) | 10.352(3) | [55] |

* $\beta$ = 96.17(3)°.

## FIGURE CAPTIONS

**Figure 1.** Crystal structures of layered *d*-metal pnictide oxides with fluorite-type [$M_2Pn_2$] layers.

**Figure 2.** Crystal structures of layered *d*-metal pnictide oxides without fluorite-type [$M_2Pn_2$] layers.

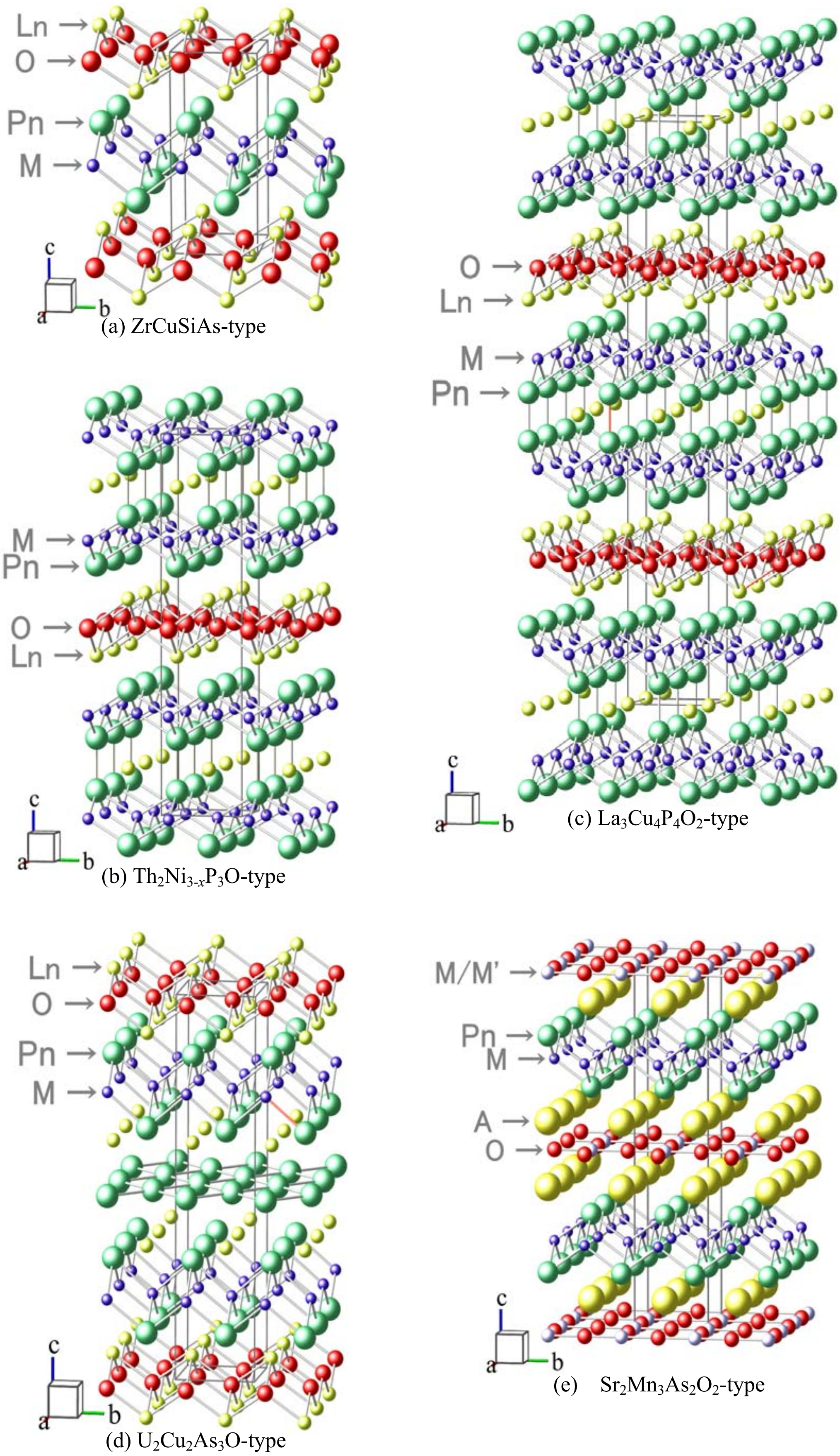

(a) ZrCuSiAs-type
(b) $Th_2Ni_{3-x}P_3O$-type
(c) $La_3Cu_4P_4O_2$-type
(d) $U_2Cu_2As_3O$-type
(e) $Sr_2Mn_3As_2O_2$-type

Figure 1

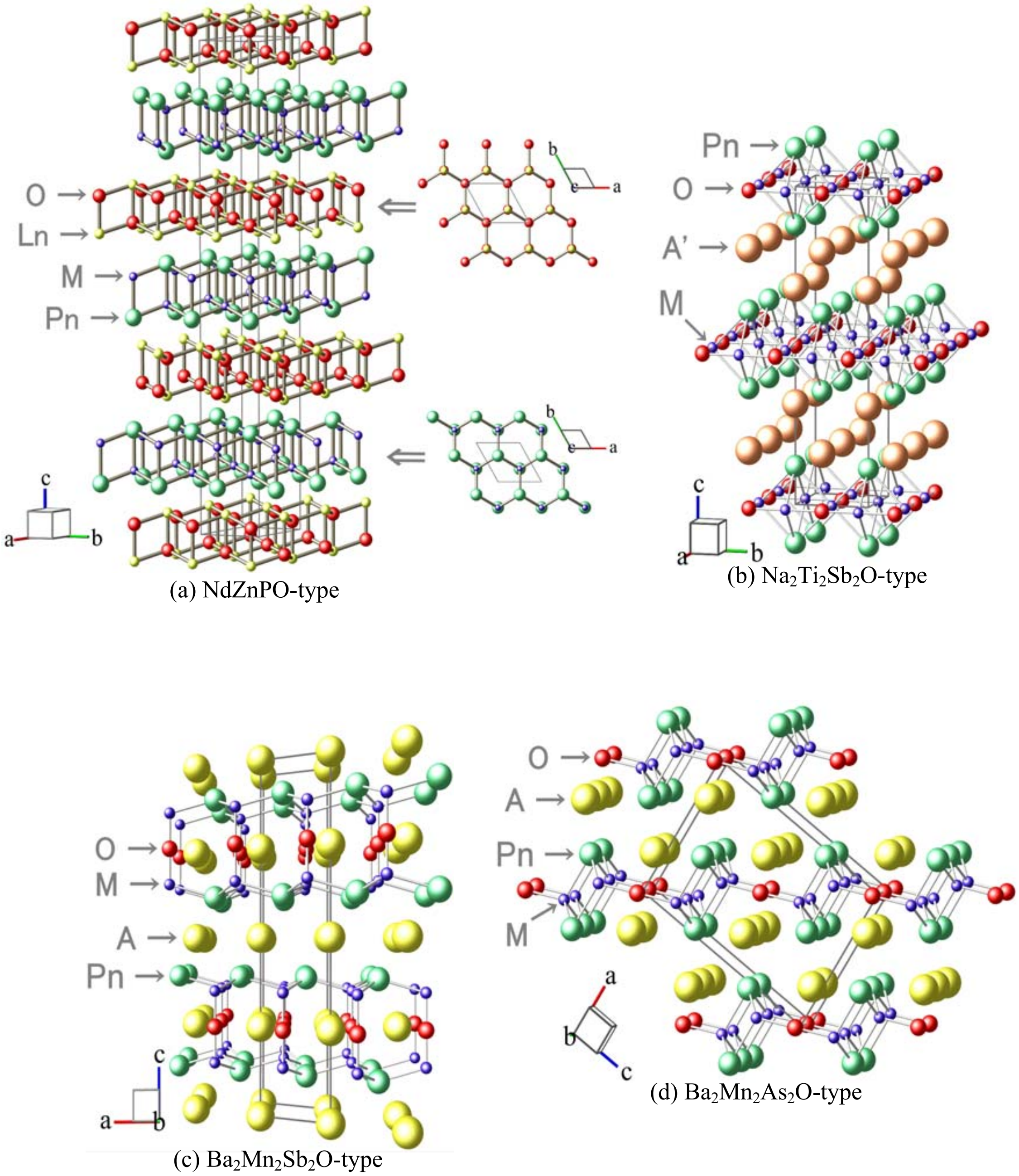

(a) NdZnPO-type

(b) $Na_2Ti_2Sb_2O$-type

(c) $Ba_2Mn_2Sb_2O$-type

(d) $Ba_2Mn_2As_2O$-type

Figure 2